\input epsf

\catcode `@=11

\def\picture #1 by #2 (#3){
\vbox{
\vskip 5mm
\vbox to #2{
\vskip #2
\special{picture #3}
\hrule width #1 height 0pt depth 0pt
\vfill
}
\vskip 5mm}}

\def\eqnarray#1{\displ@y\tabskip\centering
 \halign to \displaywidth{$\@lign\displaystyle{##{}}$\tabskip\z@skip
  &\hfil$\@lign\displaystyle{{}##{}}$\hfil\tabskip\z@skip
  &$\@lign\displaystyle{{}##}$\hfil\tabskip\centering
  &\llap{$\@lign##$}\tabskip\z@skip\crcr#1\crcr}}

\catcode `@=12
%

\newcount \nref

\def\ref {\global \advance \nref by 1 \ifnum\nref<10 \item {$ [\the\nref]~$}
\else \item{$[\the\nref]~$} \fi}

\overfullrule=0pt
\font\eightrm=cmr8

\magnification=1200
\parindent=20pt\parskip=0pt
\hsize=16.7true cm
\vsize=24.2true cm
\baselineskip=10pt
\topskip4truecm
\centerline{{\bf DEGENERATE BESS}}
\vskip5pt
\centerline{{\bf  AT FUTURE $e^+e^-$ COLLIDERS}
\footnote*{\eightrm\baselineskip=10pt
Talk presented at the IInd Rencontres du Vietnam, At the Frontiers
of the Standard Model, Hochiminh Ville, October 21-28 1995.
This work was partially supported by a grant under
the European Human Mobility Program on "Tests of Electroweak
Symmetry Breaking at Future European Colliders".
}}

\vskip15pt
\centerline{Daniele Dominici}
\centerline{Dipartimento di Fisica dell'Universit\`a and}
\centerline{Istituto Nazionale di Fisica Nucleare}
\centerline{I-50125 Firenze, Italy}
\vskip8pt
\vskip3true cm
\centerline {\bf ABSTRACT}
\vskip10pt
\noindent
The sensitivity of future $e^+e^-$ colliders to the new physics
deriving from a model of strong electroweak symmetry breaking 
is discussed. The model considered is an effective lagrangian 
description of Goldstone bosons and of 
new vector resonances degenerate in mass.
\vskip3truecm

\topskip0pt
\baselineskip=14pt
\noindent
{\bf 1. Introduction}
\vskip5pt
\def\gs{{g''}}

The model proposed in $^{1)}$ is an effective lagrangian description
of Goldstone bosons and new vector and axial vector resonances
as the possible manifestation at low energy of
the strong interacting sector. 

To build the effective low energy theory describing
Goldstones and vectors, one can use the non linear representations
of a chiral symmetry $G$ and considering (\`a la Weinberg)
the $\rho$ as the gauge field of
 the unbroken symmetry group $H$.
In a completely equivalent way one can use
the {\it hidden gauge symmetry} approach. Theories with
non linearly realized symmetry $G\to H$ can be linearly realized by
enlarging the gauge symmetry $G$ to $G\otimes H^\prime\to
H_D=diag(H\otimes H^\prime)$. $H^\prime$ is a local gauge group 
and the $\rho$
is the gauge field associated to $H^\prime$.

The BESS (Breaking Electroweak Symmetry
Strongly) model $^{2)}$ was built in this way, 
using $G=SU(2)_L\otimes SU(2)_R$,
$H=SU(2)_V$, and considering the
gauging of $SU(2)_W\otimes U(1)_Y$.

The model with vector and axial vector resonances can be built 
using the same technique.
In particular I will present the results for the 
 case
in which the new resonances are degenerate in mass (neglecting the weak
corrections).

This type of
realization corresponds to a maximal symmetry $[SU(2)\otimes SU(2)]^3$.
After gauging the standard $SU(2)_W\otimes U(1)_Y$, the model describes the
 ordinary
gauge bosons $W^\pm$, $Z$ and $\gamma$ and, in addition, 
two new triplets of massive gauge bosons, $L^\pm$, $L_3$, $R^\pm$, $R_3$,
self-interacting with gauge coupling constant $\gs$. 
These heavy
resonances, as a consequence of the chiral symmetry, are degenerate
in mass $M$ in the  $\gs\to\infty$ limit.

This degenerate model has the interesting feature of allowing 
for a strong
electroweak resonant sector at relatively low energies, while satisfying
the severe constraints from existing LEP, SLC and CDF data (see Fig. 1).

\vglue13pt
\centerline{
\epsfxsize=8truecm
\epsffile[78 263 489 700]{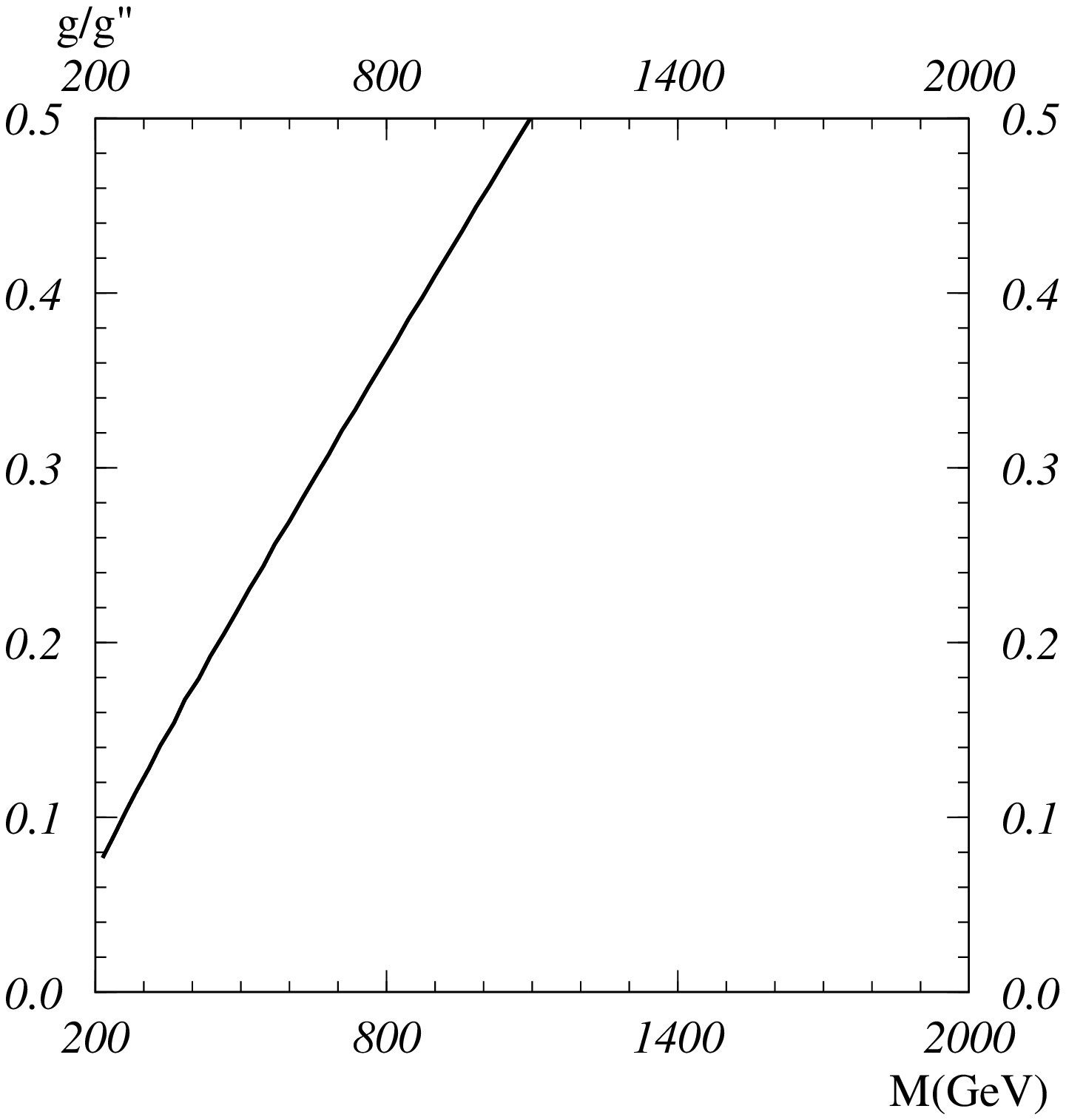}}
\vglue13pt
{\eightrm Fig.~1. 90\% C.L. contour on the plane ($M$, $g/g''$) obtained by
comparing the values of the $\epsilon$ parameters 
from the model with the experimental data from LEP.
 The allowed region is below the curve.}
\smallskip
\vskip5pt
\newcount\nfor

\def\form {\global\advance\nfor by 1 \eqno (2.\the\nfor)}
\noindent
{\bf 2. Degenerate BESS at future $e^+e^-$ colliders}
\vskip5pt
In this section I will discuss
the sensitivity of the model at LEP2 and future $e^+e^-$
linear colliders, for different options of total centre of mass energies and
luminosities. Cross-sections and asymmetries for the channel
$e^+e^-\rightarrow f^+f^-$ and $e^+e^-\rightarrow W^+W^-$ in the Standard
Model and in the degenerate BESS model at tree level have been studied.
In the fermion channel the study is based on the following observables:
the total hadronic ($\mu^+\mu^-$)
cross-sections $\sigma^h$ ($\sigma^{\mu}$),
the forward-backward and left-right
asymmetries  $A_{FB}^{e^+e^- \to \mu^+ \mu^-}$,
$A_{FB}^{e^+e^- \to {\bar b} b}$,
$A_{LR}^{e^+e^- \to \mu^+ \mu^-}$,
$A_{LR}^{e^+e^- \to h}$ and  $A_{LR}^{e^+e^- \to {\bar b} b}$.
At LEP2 we can add to the previous observables the $W$ mass measurement.

The result of this
analysis
shows that LEP2 will not improve considerably the existing limits $^{3)}$.
To improve these limits it is necessary to consider higher energy colliders.
Two options for a high energy $e^+e^-$ collider have been
studied:
$\sqrt{s}=500~GeV$ ($\sqrt{s}=1~TeV$)
with an integrated luminosity of $20 fb^{-1}$ ($80 fb^{-1}$).
In Fig. 2 a combined picture of the
 90\% C.L. contours on the plane ($M$, $g/g''$) from $e^+e^-$
at two values of $\sqrt{s}$ is shown.  The dotted
line represents the limit from the combined unpolarized observables at
$\sqrt{s}=500~GeV$ with an integrated luminosity of $20 fb^{-1}$; the
dashed line is the limit from the combined unpolarized observables
at $\sqrt{s}=1000~GeV$ with an integrated luminosity of $80 fb^{-1}$.
As expected increasing the energy of the collider and rescaling the
integrated luminosity gives stronger bounds on the
parameter space.

In conclusion
a substantial improvement with respect to the LEP bounds,
even without polarized beams, is obtained.
Considering the polarized observables one gets a further improvement on the 
limits.
The $WW$ final state does  not modify the strong limits
obtained using the fermion final state. This is  because
the degenerate model has no strong
enhancement of the $WW$ channel, present in the usual
strong electroweak models.

\vglue13pt
\centerline{
\epsfxsize=8truecm
\epsffile[78 263 489 700]{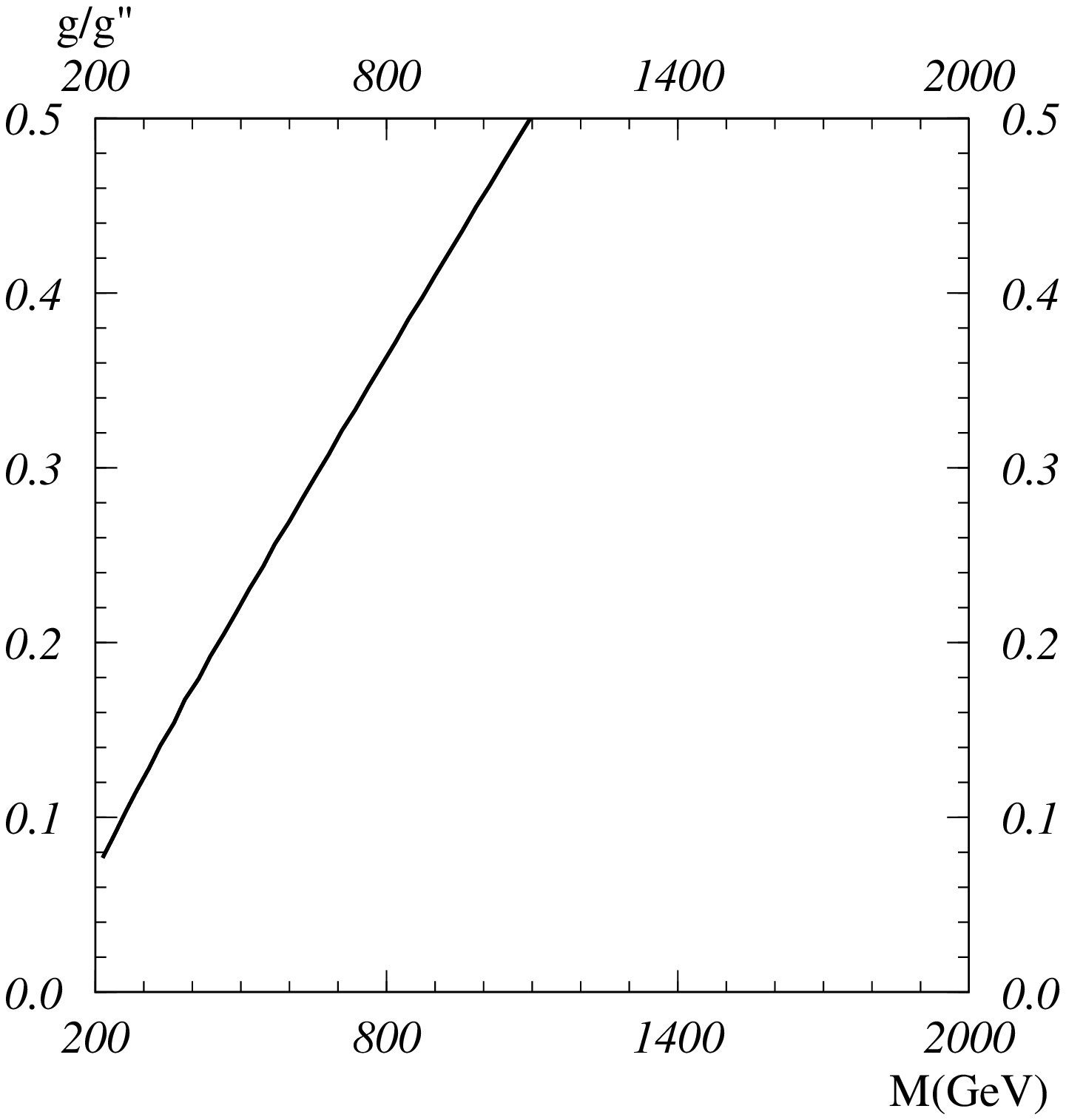}}
\vglue13pt
{\eightrm Fig.~2. 90\% C.L. contour on the plane ($M$, $g/g''$) from 
$e^+e^-$ at $\sqrt{s}=500~GeV$ with an 
integrated luminosity of $20 fb^{-1}$ and $\sqrt{s}=1000~GeV$ 
with an integrated luminosity of $80 fb^{-1}$.
Allowed regions are below the curves.}
\smallskip

\vskip20pt
\centerline{\bf References}
\vskip10pt

\baselineskip=10pt
\noindent
\ref
R. Casalbuoni, A. Deandrea, S. De Curtis, D. Dominici,  R. Gatto
and M. Grazzini, UGVA-DPT 1995/10-906,  hep-ph/9510431.
\ref
    R. Casalbuoni, S. De Curtis, D. Dominici and R. Gatto, Phys. Lett.
    {\bf B155} (1985) 95;
    Nucl. Phys. {\bf B282} (1987) 235.
\vskip3pt
\ref
R. Casalbuoni, A. Deandrea, S. De Curtis, D. Dominici,  R. Gatto
and M. Grazzini, Proceedings of the CERN Workshop on Physics
at Lep2, CERN Yellow Report, to appear in 1996.
\vskip3pt
\eject
\bye